\documentclass[a4paper,11pt]{article}
\usepackage{pos}
\usepackage{wrapfig}

\begin{document}
\title{HAGRID - High Accuracy GRB Rapid Inference with Deep learning}
\author*[a]{Merlin Kole}
\author[a]{Gilles Koziol}
\author[a]{David Droz}

\affiliation[a]{DPNC, University of Geneva, quai Ernest-Ansermet 24, 1205 Geneva, Switzerland}

\emailAdd{merlin.kole@unige.ch}
\abstract{Since their discoveries in 1967, Gamma-Ray Bursts (GRBs) continue to be one of the most researched objects in astrophysics. Multi-messenger observations are key to gaining a deeper understanding of these events. In order to facilitate such measurements, fast and accurate localization of the gamma-ray prompt emission is required. As traditional localization techniques are often time consuming or prone to significant systematic errors, here we present a novel method which can be applied on the POLAR-2 observatory. POLAR-2 is a dedicated GRB polarimeter, which will be launched towards the China Space Station (CSS) in 2025. The CSS provides POLAR-2 access to a GPU, which makes it possible and advantageous to run a Deep Learning model on it. In this work, we explore the possibility to identify GRBs in real time and to infer their location and spectra with deep learning models. Using POLAR simulations and data, a feasibility experiment was performed to implement this method on POLAR-2. Our results indicate that using this method, in combination with real time data downlinking capabilities, POLAR-2 will be able to provide accurate localization alerts within 2 minutes of the GRB onset.}
\ConferenceLogo{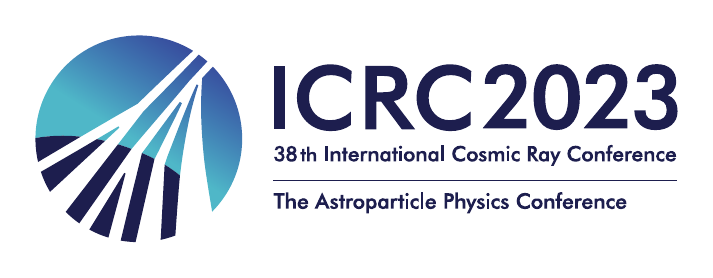}

\FullConference{%
38th International Cosmic Ray Conference (ICRC2023)\\
  26 July - 3 August, 2023\\
  Nagoya, Japan}

\maketitle

\section{Introduction}

Gamma-Ray Bursts (GRBs) are the most violent explosions in the Universe. They consist of a bright prompt emission, mostly visible in gamma-rays, followed by a longer lasting afterglow visible at lower wavelengths. The prompt emission is so bright that it can relatively easily be observed using wide field of view instruments. For typical GRBs with a brightness of $1\times10^{-5}\,\mathrm{erg/cm^2}$, an effective area of several 10's of $cm^2$ typically suffices for a detection. The afterglow, however, is significantly fainter, therefore often requiring narrow field of view instruments to achieve the signal to background required for observations. Successful afterglow measurements therefore require fast alerts from other observatories, such as wide field of view gamma-ray detectors, with an accurate location in order to catch the afterglow in time. Similarly, instruments such as MAGIC, HESS and CTA rely on such alerts to be able to redirect their telescopes to catch the GeV or TeV emission from such events.

The need for fast alerts indicates the vital role gamma-ray detectors will play in the quickly evolving multi-messenger field. This is also indicated by the large number of instruments currently in preparation with accurate localization as their primary science goal. Examples of this are MoonBEAM \cite{MoonBEAM} and HERMES \cite{HERMES}. Such detectors require not only a large sensitivity and localization capabilities, but also a system to perform fast localization calculations and to send the alerts or data to ground.

Although not designed with this as its primary science goal, the POLAR-2 detector (see ICRC proceedings \cite{Proc_Produit} for details) is competitive in its localization capabilities with for example the previously mentioned missions. The wide field of view detector, detailed in section \ref{sec:POLAR-2}, has the largest effective areas of a gamma-ray detector in space since BATSE on CGRO. This, combined with it observing half the sky, allows it to detect a large number very weak GRBs such as GRB 170817A. In addition, the segmented nature of the detector makes it sensitive to the incoming direction of the GRB. Most importantly though, the POLAR-2 detector will be placed on the China Space Station (CSS) which provides it access to on board GPU computing facilities as well as real time telemetry to Earth. Given these characteristics, POLAR-2 will be able to detect and localize a large fraction of all GRBs observed in the sky and subsequently submit detailed location and spectral information to ground within minutes. 

In order to succeed with this, reliable, fast online analysis software needs to be developed. The goal of this software is to autonomously detect GRBs in the POLAR-2 data, in real time, and subsequently perform localization and spectral analysis on the GRB data. As this software will be ran on a GPU a Deep Learning approach was investigated and tested using real data from the predecessor of POLAR-2, POLAR. Here we will present the various Deep Learning methods employed on the POLAR data, their results and how these can be employed in the future for POLAR-2.

In these proceedings we will first discus the POLAR-2 detector, particularly focusing on its characteristics which allow it to perform localization measurements. This is followed by a discussion on traditionally existing localization studies and why deep learning methods can be advantageous. Subsequently, we will present the results from the first version of the software we developed here, called the High Accuracy GRB Rapid Inference with Deep learning (HAGRID) method. We will finish with a discussion on the prospects for POLAR-2 and further studies to be performed in the near future.

\section{POLAR-2}\label{sec:POLAR-2}

POLAR-2 is a dedicated GRB polarimeter being developed by a collaboration from Switzerland, Poland, Germany and China. The mission is the successor of the POLAR mission which performed polarization measurements of 14 GRBs in 2016 and 2017. POLAR-2 will be launched towards the CSS in 2025 or 2026. From there it will observe half the sky for a duration of at least 2 years.

\begin{figure}[t]
	\begin{center}
    \includegraphics[height=.4\textwidth]{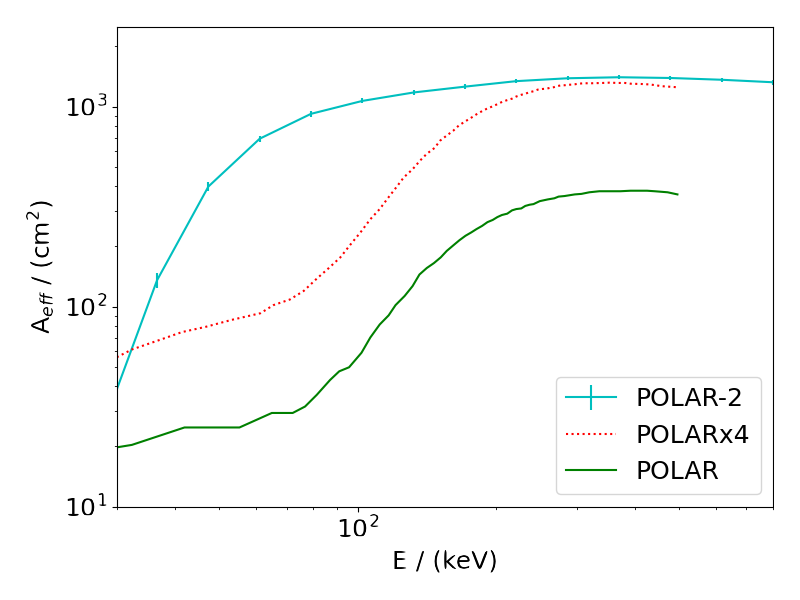}\hspace*{0.5cm}\includegraphics[height=.4\textwidth]{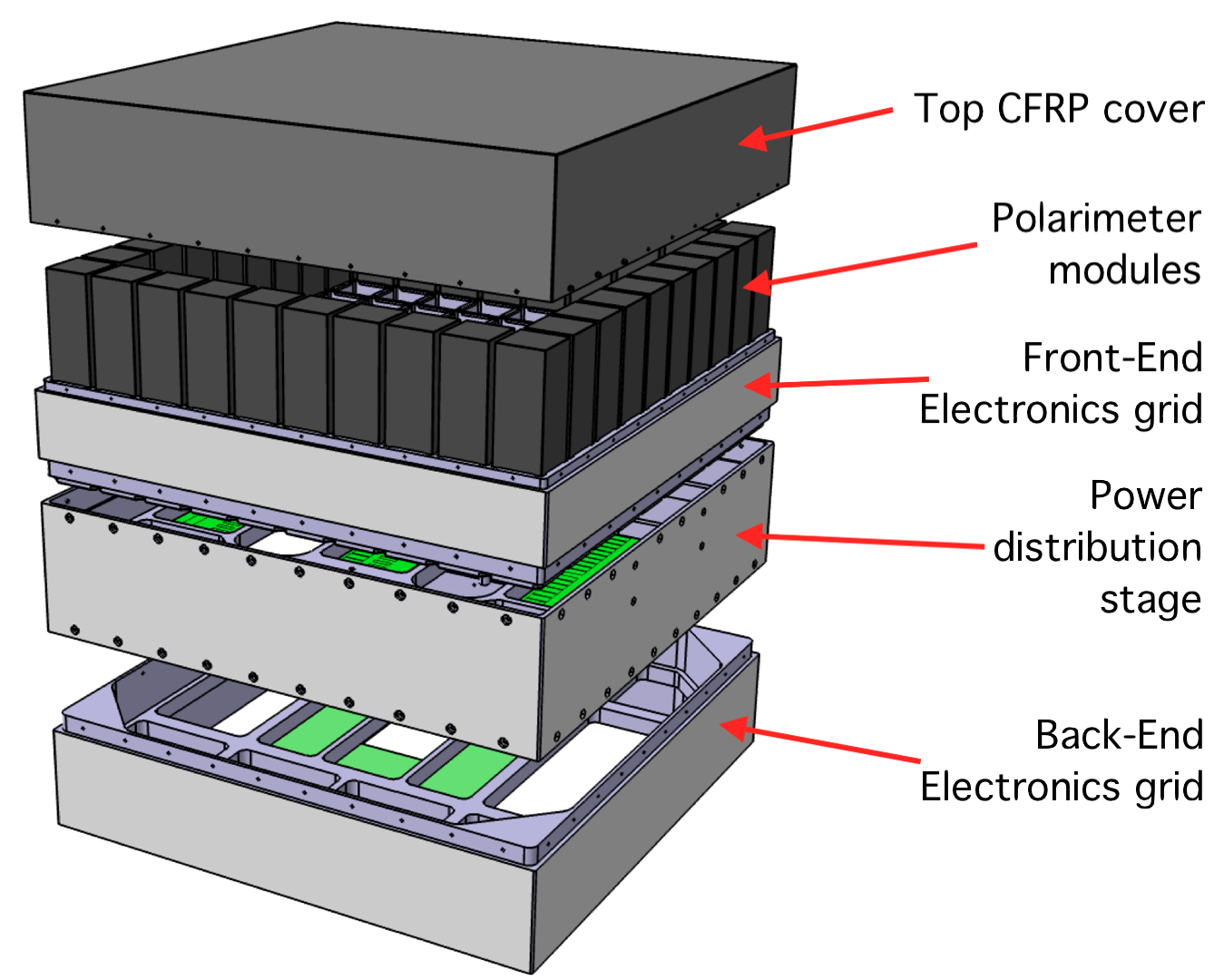}
		\caption{Left: the effective area versus initial photon energy. Green shows the effective area of POLAR, and red that of POLAR when simply increasing its size by a factor of 4. Blue shows the POLAR-2 effective area. It can be seen that POLAR-2 is significantly more sensitive at low energies thanks to the SiPM technology, Right: an exploded view of the POLAR-2 high energy polarimeter. We can see the 100 modules in black and the electronics in green}
		\label{fig:POLAR-2}
	\end{center}
\end{figure}

The POLAR-2 detector, which is presented in detail in \cite{Proc_Produit} measures the polarization of gamma-rays using a segmented scintillator detector. The detector, shown on the right of figure \ref{fig:POLAR-2}, therefore consists of an array of 6400 scintillator bars with dimensions of $5.9\times5.9\times125\,\mathrm{mm^3}$ each of which is readout using a Hamamatsu S13 MPPC. The plastic scintillator bars are optimized for the purpose of polarimetry, meaning that they have a low atomic number to optimize the cross section for Compton scattering (required to measure the polarization) in the 20-800 keV energy range in which POLAR-2 will be sensitive. 

In order to measure polarization, an incoming gamma-ray needs to Compton scatter in one plastic scintillator and subsequently interact in a second one. This allows one to measure the azimuthal Compton scattering angle which is correlated to the intrinsic polarization. Although POLAR-2 is optimized to measure this, the efficiency for such interactions can be imagined to be relative small. This as, opposed to the preferred interaction mechanism, some gamma-rays can undergo photo-absorption only, scatter out of the detector after their first Compton scattering interaction, or scatter several times in the array. As a result, only about $10\%$ of all the incoming photons can be used for gamma-ray polarimetry, making such measurements statistically starved.

To compensate for this POLAR-2 will have a large surface area of around $600\times600\,\mathrm{cm^2}$. Its effective area for localization studies exceeds $2000\,\mathrm{cm^2}$. When only taking photons into account which can be used for polarimetry the effective area is around $1000\,\mathrm{cm^2}$. This is compared with that of POLAR on the left of figure \ref{fig:POLAR-2}. In addition, in order to observe as many GRBs as possible the instrument has a field of view of half the sky. This large field of view combined with the large effective area, and the improved signal to background compared to POLAR, allows POLAR-2 to observe GRBs with fluences as low as $10^{-8}\,\mathrm{erg/cm^2}$.

Furthermore, the segmented nature of the detector makes it sensitive to the incoming direction of GRBs. The sensitivity is similar to that used in Fermi-GBM, where the dependence of the effective area of the various sub-detectors on the incoming direction of the GRB can be used to infer the location of the GRB in the sky. This method was tested for POLAR in the past and presented in \cite{Wang} where traditional localization methods were used to find that GRBs could be localized to within several degrees. The larger effective area of POLAR-2, combined with the larger number of sub-detectors (6400 compared to 1600 for POLAR) will likely increase this precision.

Finally, POLAR-2 will be placed on the CSS in 2025. As such, its data can be analyzed in near real time using a GPU (an nvidia TX2). The results of the GPU analysis can subsequently be submitted to ground using either the Beidou satellite system or using the CSS real time telemetry. In case the POLAR-2 data can be analyzed autonomously this will allow for alerts to be send to ground within 2 minutes of the onset of a GRB, thereby allowing instruments such as CTA, MAGIC and HESS to repoint to the GRB location to capture the high energy emission.

\section{Localization Methods}

\begin{wrapfigure}{r}{0.5\textwidth}
\begin{center}
\includegraphics[height=6.0cm]{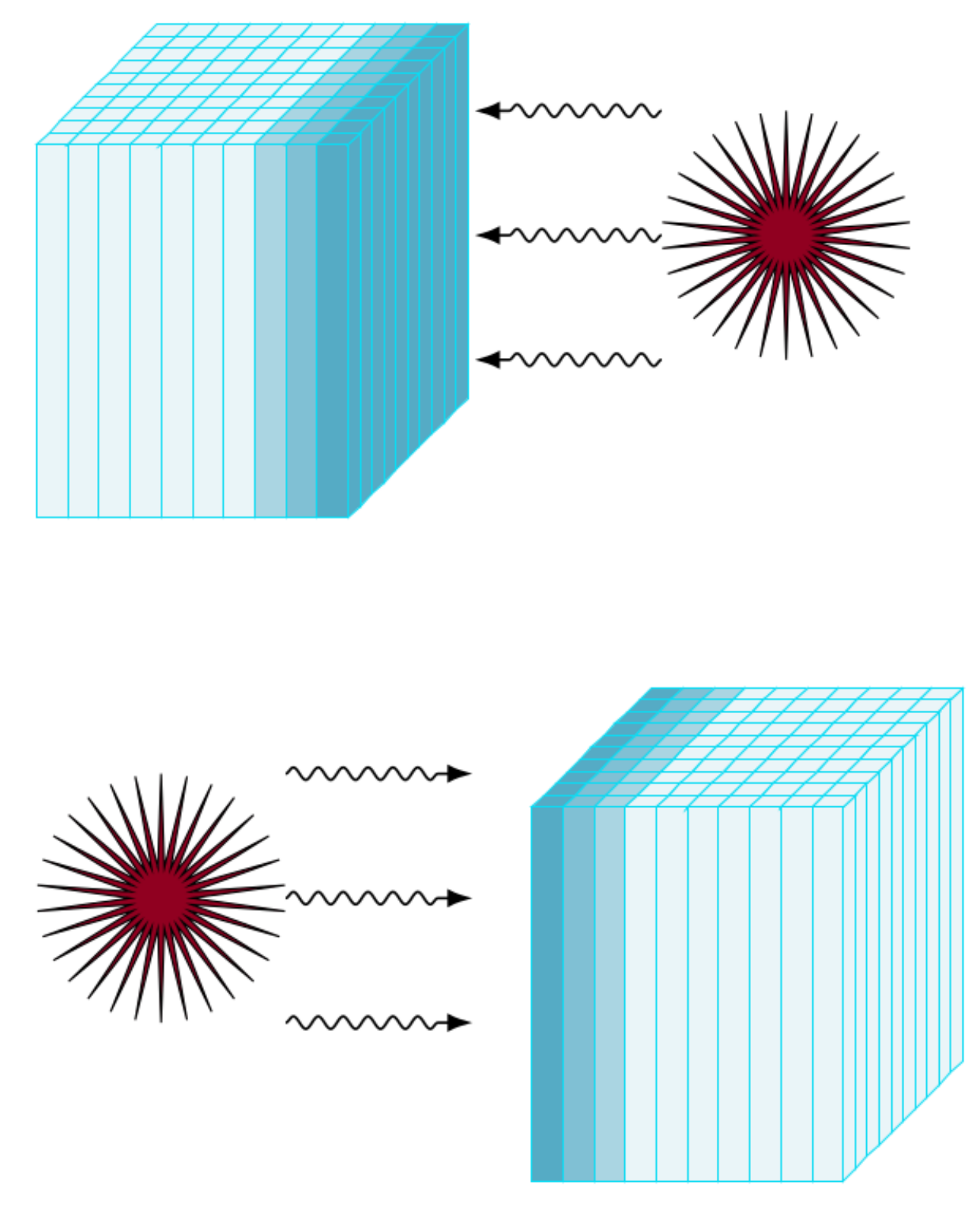}
\caption{Illustration of the idea behind localization measurements of a GRB using a segmented detector. The number of photons interaction in the various detector elements (illustrated by the darkness of the elements) depends on the incoming direction of the GRB.}
		\label{fig:loc_idea}
  \end{center}
\end{wrapfigure}

The POLAR and POLAR-2 GRB localization method makes use of the dependence of the effective area of the various scintillator bars on the incoming direction of the photons. For example, while a GRB occurring at zenith will result in an equal effective area for all 6400 bars, a GRB coming from the side will result in the scintillator bars on the side to have a significantly larger effective area than those on the opposite side as these are shadowed by other materials. This idea is illustrated in figure \ref{fig:loc_idea}. This method is similar to that employed for example for Fermi-GBM or BATSE.

A very simple analysis method would be to compare the relative number of observed photons in each detector element with those predicted using Monte Carlo (MC) simulations. Comparing the observed relative rates vs those predicted through MC simulations for all possible incoming angles allows to identify the most likely incoming angle of the GRB. A clear drawback of this method is that this relative number of counts not only depends on the incoming direction of the GRB, but also on the spectral shape of the GRB. To somewhat mitigate this, one can expand the array of MC simulation results by performing these simulations for 3 different typical GRB spectral types (hard, medium and soft), thereby producing a 3d matrix (with dimensions of spectral shape, $\theta$ and $\phi$) against which the observed data can be compared using, for example, a $\chi^2$ fit. This method, which is fast and relatively easy to perform, is employed on Fermi-GBM data \cite{GBM} as well as on the POLAR data in \cite{Wang}. 

A clear downside of this method is that the real spectral shape of the observed GRB will not perfectly match that of one of the 3 simulated GRBs, thereby inducing systematic errors in the final location result \cite{BALROG}. A more complex and time consuming version of this method uses the best fit location to perform a spectral fit, the result of which is subsequently used to perform a second localization fit using a simulated array produced for a spectral shape similar to the best fitted spectral fit. Although this method provides a more accurate result, the issue with the systematic errors remains.

A more sophisticated method which overcomes this issue is the BALROG method \cite{BALROG}. Opposed to the method described above, the BALROG analysis performs a joint localization and spectral fit on the data. By fitting both at the same time the best spectral and location are found together, thereby mitigating issues with systematic effects. The downside of this method is that it is significantly more computationally demanding. Given advances in computing power this issue is not significant when performing on ground analysis, making this method preferable for this, however, for performing real time analysis on a space station this issue is a significant downside.

Although the simple $\chi^2$ method can be employed in space, it is far from ideal and does not make use of the equipment available to POLAR-2. Here we therefore present a third option which makes use of Deep Learning. This method can in principle be used to perform fast, accurate analysis where systematic errors are automatically included in the results. The further advantage is that the localization calculation can be performed very quickly, especially on a GPU. The training of the model can be performed on ground using real data during the mission and updated models can be uploaded to the CSS and applied in real time to data.

\section{HAGRID Results}

The first job of the HAGRID software is to autonomously detect GRBs within the POLAR-2 light curve. For this method a Long Short-Term Memory model was trained using both POLAR background data and simulated GRBs produced using the POLAR simulation software. In total over 150'000 artificial GRBs were produced using real background data from POLAR, where the rate vs time was picked based on a polynomial shape. Artificial GRBs (with random spectral shapes and incoming directions) were subsequently placed on top of these background light curves. The energy and interactions of these GRBs were produced using MC simulations while the light curves were produced using part of the CosmoGRB package \footnote{\url{https://github.com/grburgess/cosmogrb}}. The input data for the model to train on consisted of both the rates of the various sub-detectors, as well as the measured energy spectra. This allows the model to not only detect transient events based on the increasing rates, as is often done by eye, but to also make use of the changes in the spectral shapes as well as the relative differences in the rates in the various sub-detectors.

\begin{figure}[t]
\includegraphics[height=4.4cm]{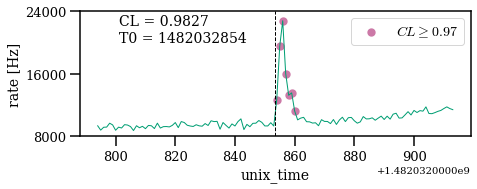}
\caption{GRB 161218A as identified correctly by HAGRID in the POLAR data.}
		\label{fig:fake_GRBs}
\end{figure}

After training on the large simulation sample the produced model was tested on several days of real POLAR data. Although various minor issues were found (mostly related to the POLAR detector being switched on and off around the SAA which was recognized as a GRB) the model recognized all real GRBs known to be in the studied data. An  example of this can be seen in figure \ref{fig:fake_GRBs}. In addition, HAGRID identified unknown GRB candidates while the false positive rate (with exception to SAA induced events which can easily be filtered out) remained negligible.

Subsequent to correctly detecting a GRB, HAGRID is designed to perform spectral and localization analysis on the GRB data. As the GRB detection algorithm currently studies the GRB lightcurve in 1 second bins, the localization and spectral studies commence 1 second after the onset of the GRB. If the GRB is longer the analysis will be performed each subsequent second, on the total accumulated data, until the GRB stops.

For the spectral analysis again a large sample of simulated GRBs (using the POLAR simulation software) was used to train the model. In these simulations, again the incoming angle of the GRB was randomized, while the spectral shape was simulated as being a Band function. The Band function parameters ($\alpha, \beta$ and $E_{peak}$) were simulated from distributions as measured by Fermi-GBM and BATSE. 

\begin{figure}[h]
         \centering
         \includegraphics[width=8cm]{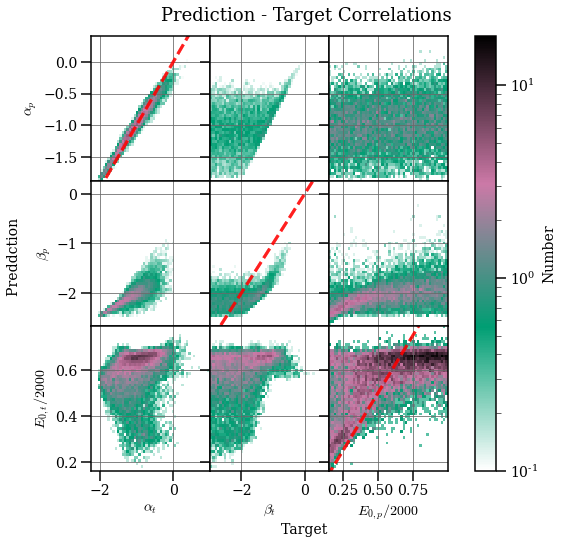}
         \caption{Correlation between target and predictions of model $M_{sep}$. This model predicts the $\alpha$ parameter quite accurately, whereas the $\beta$ and $E_0$ parameters are harder to predict. Note however that the model always predicts $\alpha>\beta$}
         \label{fig:spec}
\end{figure}

For the model, many options were studied using a toy MC analysis. It was found that Fully Connected Neural Networks making use of loss functions with mean squared error and with the three output parameters being treated independently performed best. The results on the POLAR simulation data are illustrated in figure \ref{fig:spec} where the input parameters for the three spectral parameters are shown as the Target on the x-axis and the reconstructed ones on the y-axis. As expected with the energy range of POLAR (50-500 keV) the reconstruction of the $\alpha$ parameters is successful, while both $E_{peak}$ and $\beta$ are more difficult to reproduce. In detailed follow-up studies it was confirmed that the issues in this reproduction are due to the lack of photons at energies above the $E_{peak}$ value. As can be seen in figure \ref{fig:spec} for low values of $E_{peak}$ a somewhat better reproduction can be performed. 

\begin{figure}[h]
         \centering
         \includegraphics[width=8cm]{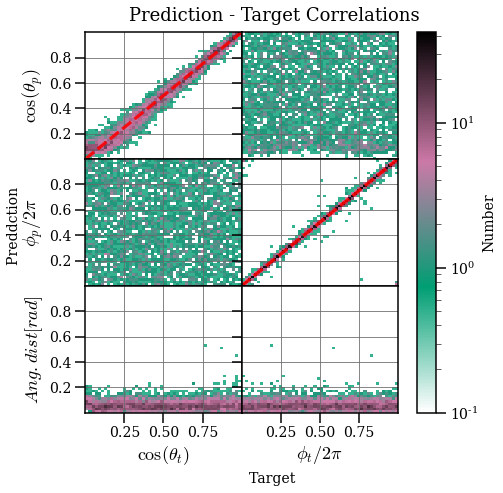}
         \caption{Correlation between the target and predictions for the deeper convolutional neural network. The model is performing well, except for events that are coming from the equator. The $\phi$ predictions are good since very few outliers are to be reported.}
         \label{fig:loc}
     \end{figure}

Finally, and most importantly, a localization model is applied to the data in parallel to the spectral model. For the localization model a vast array of models was tested both on toy MC and on real POLAR data. It was concluded that a deep 2d convolutional neural network performed best on the data. Furthermore, the model was found to perform best when trained for outputting $\theta$, $\cos(\theta)$ and $\sin(\phi)$. A result of the correlation between the target and reconstructed location angles for a deep convolutional neural network is shown in figure \ref{fig:loc}. The median localization error on this data set was found to be $3.6^\circ$. It thereby performs better than the traditional methods used on the POLAR data in \cite{Wang}.

\section{Conclusions and Discussion}

The preliminary results of the performance of HAGRID show great promise. The model, which is currently only trained using POLAR data as no POLAR-2 data is yet available, already performs well for GRB detection, while spectral and localization studies also show good results. The latter two can however be greatly improved with further studies.

Firstly, it was found that the current training sample of 150'000 artificial GRBs is not yet sufficient for accurate training. A study of the performance of the models versus the number of training GRBs shows that increasing the number will increase both the localization as well as the spectral performance. This issue can be easily overcome by running more simulations which will be performed during the coming months.

Furthermore, it was found that the normalization of the data sets, meaning normalization of the various input parameters in order to optimize the training, can be improved. This requires further studies during the coming months.

Finally, the training will start to be performed on the POLAR-2 MC data instead of the on the POLAR MC. This will indicate whether the larger effective area of POLAR-2 and its larger number of detector elements can improve the localization as is expected. Furthermore, the increased energy range of POLAR-2 (up to 800 keV compared to 500 keV of POLAR), will allow for better spectral reconstruction especially of the $E_{peak}$ and $\beta$ parameters.

Overall we currently expect to employ the HAGRID method on the CSS after the launch of POLAR-2. Initially we will need to train the model using real POLAR-2 data, and using flight data verified MC simulations of GRBs. Once the results are optimized we anticipate to run HAGRID autonomously on POLAR-2 data several months after launch with the aim of producing alerts with degree level localization precision within 2 minutes of the GRB onset.

\section{References}

\bibliographystyle{JHEP}
\bibliography{my-bib-database}
\end{document}